\documentclass{iopjournal}

\usepackage{bm}
\usepackage{braket}
\usepackage{graphicx}
\usepackage{hyperref}

\newcommand{\dd}{\mathrm{d}}

\usepackage{bbold}
\usepackage{color}

\usepackage[]{amsmath}
\usepackage{braket}
\usepackage{amsfonts,amsmath}
\usepackage{epsfig,amsmath}
\usepackage{graphicx}
\usepackage{dcolumn}
\usepackage{bm}
\usepackage{bbold}
\newcommand{\beq}{\begin{equation}}
\newcommand{\eeq}{\end{equation}}
\newcommand{\beqa}{\begin{eqnarray}}
\newcommand{\eeqa}{\end{eqnarray}}

\newcommand{\ra}{\rangle}

\newcommand{\proj}[1]{\ket{#1}\!\bra{#1}}
\newcommand{\WU}{\mathcal W}

\begin{document}

\articletype{Research Paper} 

\title{When Malus' Law Meets the Wigner--Bell Inequality: Violation by Nonseparable Classical Light}

\author{Xiao-Feng Qian$^{1,*}$\orcid{0000-0000-0000-0000}}

\affil{$^1$Department of Physics, and Center for Quantum Science and Engineering, Stevens Institute of Technology, Hoboken, NJ 07030, USA}

\affil{$^*$xqian6@stevens.edu}


\keywords{Bell inequalities; Wigner inequality; classical optical nonseparability; joint probabilities; measurement compatibility; quantum-classical boundary}

\begin{abstract}
Bell-type inequalities are commonly associated with quantum nonlocality, yet their mathematical structure also reflects more general constraints on the joint probability distributions assigned to measurements performed in different contexts. Here we examine this distinction through the Wigner--Bell inequality using a classical optical field with two nonseparable degrees of freedom. Starting from a subsystem response-factorization assumption, we derive a Wigner-type probability bound requiring the pairwise measurement statistics to be compatible with a single nonnegative global joint distribution. For a classically nonseparable optical field, the corresponding joint probabilities are represented directly by normalized joint intensities and obey Malus' law. We show that these Malus-law probabilities violate the Wigner--Bell bound over a finite range of analyzer settings. The violation does not imply spatial nonlocality or a failure of classical electrodynamics. Rather, it demonstrates the incompatibility of context-dependent optical projection statistics with the global probability assignment underlying the Wigner construction. We further derive an extended Wigner-Bell bound for general imperfect correlations and propose an interferometric experiment for reconstructing the required joint probabilities from intensity measurements. These results provide a transparent optical realization of the probabilistic structure behind Bell-type inequalities and help distinguish consequences of Hilbert-space structure, nonseparability, and measurement context from those specifically associated with quantum nonlocality.
\end{abstract}

\section{Introduction}
\label{sec:introduction}

The phrase ``Bell inequality'' encompasses several logically related but
operationally distinct constraints. Bell's original three-setting argument combined locality with the perfect anticorrelation of an Einstein-Podolsky-Rosen pair
\cite{Bell1964}. The Clauser-Horne-Shimony-Holt (CHSH) formulation removed
this ideal-correlation requirement by combining correlations from four
setting pairs \cite{CHSH1969}. Wigner recast the Bell obstruction in terms of
joint events and probability sets \cite{Wigner1970}, making explicit the requirement that pairwise distributions measured in mutually exclusive contexts be marginals
of a single nonnegative global joint distribution
\cite{Wigner1970,Fine1982,Fine1982JMP,Pitowsky1989,Eberly2002, Abramsky2011}. The Clauser-Horne (CH) inequality subsequently provided a probability-based formulation that, unlike Wigner's original inequality, requires neither perfect correlation nor perfect anticorrelation and explicitly includes both
coincidence and single-detection probabilities \cite{CH1974}. For complete binary-outcome measurements, CH and CHSH are equivalent representations of the same local constraint, although CH is more directly suited to experiments
with inefficient photodetection \cite{Fine1982, Eberhard1993}.

A classical optical field with two nonseparable degrees of freedom has the
same tensor-product amplitude structure as a bipartite pure quantum state \cite{Spreeuw1998,Lee2002,Luis2009,Borges2010,Goldin2010,Simon2010,
Qian2011,Kagalwala2013,Pereira2014,Toppel2014,de2014relationship,
Aiello2015,Berg-Johansen2015,Qian2015,Rafsanjani2015,Stoklasa2015,
Li2015,Chekhova2015,Svozilik2015,Qian2016,Silva2016,Balthazar2016,
Eberly2016,Ndagano2017,Qian2017,Eberly2017,Qian2018,Zela2018,Korolkova2019,Forbes2019,
Goldberg2021,Hu2021,Nape2022,Ares2023,Alonso2023,Khalid2025,
hu2025robust,Marletto2025,Abouraddy2026,Braga2026,Laatikainen2026}. Polarization, for example, can be nonseparable from a spatial \cite{Borges2010,Kagalwala2013, Toppel2014,Aiello2015,Berg-Johansen2015,Ndagano2017,Nape2022,Laatikainen2026}, temporal \cite{Qian2011,de2014relationship}, path \cite{Spreeuw1998,Qian2016,Silva2016, Eberly2017, Qian2018}, or statistical-amplitude mode \cite{Simon2010,Qian2015}.
Particularly, such fields have reproduced quantum-like Bell--CHSH
correlations \cite{Spreeuw1998, Borges2010, Goldin2010,Kagalwala2013, Qian2015, Stoklasa2015, Ares2023}, including experimental violations of the CHSH bound by
classical light \cite{Borges2010,Kagalwala2013,Qian2015,Stoklasa2015}. These correspondences do not transform the internal degrees of freedom of a single beam into two spacelike-separated particles. Instead,
they provide a controlled setting in which to distinguish phenomena arising
from Hilbert-space structure, coherence, and measurement context from those
requiring quantum particle statistics or spatial locality.

The Wigner formulation \cite{Wigner1970} of the Bell inequality has a particularly natural connection to classical
optical fields because its joint probabilities can be represented directly
by normalized joint intensities. Malus' law supplies their squared
trigonometric dependence, allowing the conflict between a global
probability-space constraint and the measured optical projections to be
displayed without first constructing correlation coefficients \cite{Eberly2002}. The resulting
violation reveals an incompatibility between a triangle-like classical event
relation and the Malus-law intensities obtained in different measurement
contexts. It represents neither a failure of Malus' law nor a breakdown of
classical electrodynamics, but provides a transparent perspective from which
to revisit the probabilistic assumptions underlying Bell-type violations.

Here we develop a direct route from the Bell and CHSH correlation
inequalities to a Wigner--Bell probability inequality for a nonseparable
classical optical field. Because the two optical subsystems are internal
degrees of freedom rather than spatially separated particles, spatial
locality is replaced by a precisely defined \emph{subsystem
response-factorization} assumption. This assumption should not be confused
with unconditional statistical independence: the observed degrees of freedom
are correlated and, for the field considered here, nonseparable. We derive
the resulting Wigner bound, demonstrate its conflict with Malus-law joint
intensities of light, and propose an experiment inspired by the joint-projection interferometer of
Ref.~\cite{Qian2015}. We also derive an imperfect-correlation correction so
that the proposed test does not depend on ideal state preparation. Together, these results illuminate assumptions that are often left implicit in Bell-type arguments.

This article is organized as follows. Section 2 briefly reviews the original Bell inequality, the CHSH inequality, and the Wigner–Bell inequality. Section 3 derives the Wigner–Bell inequality for two internal degrees of freedom of a classical light beam. Section 4 demonstrates its violation with a maximally entangled classical light field. Section 5 extends the Wigner–Bell inequality to the experimentally relevant case of imperfect equal-setting correlations. Section 6 proposes an experiment to test the violation of both the original and extended Wigner–Bell inequalities using classically entangled thermal light. Section 7 discusses the physical interpretation of these violations in the context of classically entangled light. Finally, Section 8 concludes the article.

This article is dedicated to the memory of Joseph H. Eberly, whose work repeatedly employed conceptually transparent optical models to sharpen
questions concerning entanglement, coherence, and the quantum--classical
boundary.

\section{From Bell correlations to Wigner probabilities}
\label{sec:bell-to-wigner}

\subsection{Bell's original inequality and perfect anticorrelation}

Bell's original scenario considers two separated stations with settings $a$ and $b$ and binary results
$A(a,\lambda),B(b,\lambda)\in\{-1,+1\}$.  A deterministic notation causes no
loss of generality for the inequalities below because any local stochastic
random seed may be included in the hidden variable $\lambda$. Measurement independence requires the normalized hidden-variable distribution $\rho(\lambda)$, with $\int\dd\lambda\rho(\lambda)=1$, to be independent of the measurement settings. Locality
requires the measurement outcome at either station, conditioned on $\lambda$, to be independent of the setting and outcome at the other station, i.e., 
\begin{equation}
 A(a,b,B,\lambda)=A(a,\lambda),\qquad
 B(a,b,A,\lambda)=B(b,\lambda).
 \label{eq:bell-locality}
\end{equation}
Bell's 1964 construction additionally invokes perfect anticorrelation for
equal settings $a$,
\begin{equation}
 B(a,\lambda)=-A(a,\lambda).
 \label{eq:anticorrelation}
\end{equation}
The correlation is then defined as
\begin{equation}
E(a,b)=\int\dd\lambda\,\rho(\lambda)A(a,\lambda)B(b,\lambda).
\end{equation}
Using Eq.~\eqref{eq:anticorrelation} and the binary identity
$|A(b,\lambda)-A(c,\lambda)|=1-A(b,\lambda)A(c,\lambda)$ gives the three-setting Bell inequality
\begin{equation}
|E(a,b)-E(a,c)|\leq 1+E(b,c).
 \label{eq:original-bell}
\end{equation}
The exact sign convention changes if perfect correlation is used instead of
anticorrelation, but the logical role of the perfect equal-setting relation is
unchanged.

It is common to summarize the assumptions as ``locality'' and ``realism''.  More
explicitly, realism here means that the candidate model assigns an outcome, via $\lambda$, to every relevant setting, including settings not selected in a particular run. Locality assumes results from one station to be independent of the outcome and settings of the other, while measurement independence allow results from different setting pairs to be represented with the same $\lambda$ and the same distribution
$\rho(\lambda)$. Perfect anticorrelation then identifies one party's equal-setting result with the negative of the other's.

\subsection{CHSH without perfect anticorrelation}

CHSH uses two independent settings, $a,a'$, for the first station and two, $b,b'$, for the second. For a given $\lambda$, a special combination of joint events are constructed
\begin{align}
 S(\lambda)={}&A(a,\lambda)B(b,\lambda)
 +A(a,\lambda)B(b',\lambda) \nonumber\\
 &+A(a',\lambda)B(b,\lambda)
 -A(a',\lambda)B(b',\lambda).
\end{align}
Because each response, $A(a,\lambda),B(b,\lambda)$, etc., is $\pm1$, averaging with the normalized distribution $\rho(\lambda)$ yields
\begin{equation}
 \left|E(a,b)+E(a,b')+E(a',b)-E(a',b')\right|\leq2.
 \label{eq:chsh}
\end{equation}
No perfect equal-setting correlation is required.  CHSH therefore provides a
more robust experimental inequality, at the cost of using four setting pairs
and a correlation combination rather than the three joint-event probabilities
used below in the Wigner scenario.

\subsection{Wigner's global joint-probability construction}

The Wigner formulation makes explicit the assumption of a common global
probability space \cite{Wigner1970}. The potential outcomes associated with the two
subsystems can be defined similarly as the Bell and CHSH treatment, i.e., $A(a,\lambda), B(b,\lambda) \in\{+1,-1\}$, which denote the outcomes assigned to settings $a$ and $b$, respectively, for a given hidden variable
$\lambda$. Although only one setting is selected for each subsystem in an
individual experimental run, the hidden-variable model assigns outcomes to
all relevant settings simultaneously.

It is then assumed that these outcomes possess a single nonnegative global joint distribution,
\begin{equation}
p(A_\alpha,A_\beta,B_\beta,B_\gamma)\geq0,
 \qquad
\sum_{A_\alpha,A_\beta,B_\beta,B_\gamma=\pm1} p(A_\alpha,A_\beta,B_\beta,B_\gamma)=1,
 \label{eq:globaldistribution}
\end{equation}
where, for compactness, it is defined $A_a\equiv A(a,\lambda), B_b\equiv B(b,\lambda)$ with $a=\alpha,\beta, b=\beta, \gamma$. The pairwise probabilities measured for different choices of settings are
required to be marginals of this same global distribution. For example,
\begin{equation}
 P_{++}(\alpha,\gamma)
 =
 \sum_{A_\beta,B_\beta=\pm1}
 p(+,A_\beta,B_\beta,+).
 \label{eq:pairwise-marginal}
\end{equation}

For the correlated convention adopted here, perfect equal-setting correlation, within the global hidden-variable construction, restricts the
support of the distribution to outcomes satisfying $A(\beta,\lambda)=B(\beta,\lambda)$, for every $\lambda$. It then leads to the vanishing of the two joint probabilities
\begin{equation}
 P_{+-}(\beta,\beta)=P_{-+}(\beta,\beta)=0.
 \label{eq:perfect-correlation}
\end{equation}

For a demonstration of obtaining the Wigner-Bell inequality, we start with the consideration of the event $\{A_\beta=+1,B_\gamma=+1\}$. For every member of this event, the potential outcome $A_\alpha$ must be
either $+1$ or $-1$. If $A_\alpha=+1$, the event belongs to
$\{A_\alpha=+1,B_\gamma=+1\}$. If $A_\alpha=-1$, perfect equal-setting
correlation gives $B_\beta=A_\beta=+1$, so the event belongs to
$\{A_\alpha=-1,B_\beta=+1\}$. Therefore, it follows straightforwardly 
\beq
 \{A_\beta=+1,B_\gamma=+1\}
 \subseteq{}
 \{A_\alpha=+1,B_\gamma=+1\}
\cup
 \{A_\alpha=-1,B_\beta=+1\}.
 \label{eq:event-inclusion}
\eeq
Taking the probabilities of these events and using the nonnegativity of the
global distribution yields the Wigner--Bell inequality
\begin{equation}
 P_{++}(\alpha,\gamma)
 +
 P_{-+}(\alpha,\beta)
 \geq
 P_{++}(\beta,\gamma).
 \label{eq:wigner-bound}
\end{equation}

The same inequality can also be derived directly at the probability level by
expressing each of the three pairwise joint probabilities,
$P_{++}(\alpha,\gamma)$, $P_{-+}(\alpha,\beta)$, and
$P_{++}(\beta,\gamma)$, as a marginal of a common four-variable global joint
distribution \eqref{eq:globaldistribution}. Expanding these marginals over the outcomes of the two unobserved variables makes the origin of the Wigner-Bell inequality explicit. That is, after cancellation of the common terms, the difference
between the left-hand and right-hand sides consists entirely of nonnegative
four-variable joint-probability terms.

Thus, the Wigner-Bell inequality is not merely a trigonometric relation. It is a Boole-type consistency condition \cite{Boole1862, Pitowsky1994} on selected marginals of the global
distribution in Eq.~\eqref{eq:globaldistribution}. Its violation means that
the pairwise distributions measured in the three different contexts
$(\alpha,\gamma)$, $(\alpha,\beta)$, and $(\beta,\gamma)$ cannot all be
reproduced as marginals of a single nonnegative joint distribution.

Bell, CHSH, and Wigner inequalities are not fundamentally different theories.
Rather, they express complementary consequences of a common class of
hidden-variable models. Wigner's formulation is especially transparent
because it makes explicit both the counterfactual assignment of outcomes to
alternative settings and the requirement that the measured pairwise
probabilities be marginals of a single nonnegative global distribution. This
probability-based formulation is particularly well suited to nonseparable
classical light, for which the relevant joint probabilities are represented
directly by normalized joint intensities whose angular dependence is governed
by Malus' law.

\section{Translation to two internal optical degrees of freedom}
\label{sec:optical-translation}

\subsection{Nonseparable classical field}
Classical tensor fields  are regarded to share the structure with quantum states that host the inter-space nonseparability \cite{Eberly2016,Korolkova2019,Forbes2019, Goldberg2021,Alonso2023}. A general optical beam can be expressed as 
\beq
\vec{\mathbf E}(r_{\perp},t)=\hat{\mathbf x}E_x(r_{\perp},t)+\hat{\mathbf y}E_y(r_{\perp},t),
\eeq
where $\hat{\mathbf x}, \hat{\mathbf y}$ are unit polarization directions and $E_x(r_{\perp},t), E_y(r_{\perp},t)$ are corresponding amplitudes with $r_{\perp}$ representing the transverse spatial dependence. A normalized beam in the Schmidt form \cite{Qian2011, Qian2015} can be described as
\begin{equation}
 \ket e=\frac{\ket{\mathbf E}}{\sqrt I}
 =\kappa_1\ket{\mathbf u_1}\ket{f_1}
 +\kappa_2\ket{\mathbf u_2}\ket{f_2},
 \quad \kappa_j\geq0,\quad \kappa_1^2+\kappa_2^2=1.
 \label{eq:schmidt}
\end{equation}
Here the Dirac notation is to indicate the vector nature of these degrees of freedom. The vectors $\ket{\mathbf u_j}$ span a two-dimensional polarization space $U$, and $\ket{f_j}$ span the occupied subspace of a function space $F$,
such as a pair of spatial or temporal modes, statistical amplitudes, etc. They obey the orthonormal condition
$\braket{\mathbf u_i|\mathbf u_j}=\braket{f_i|f_j}=\delta_{ij}$. $I=\braket{\mathbf E|\mathbf E}$ is the total intensity under the relevant spatial, temporal, or ensemble inner
product.

Now we introduce the real basis rotations
\beqa
 \ket{\mathbf u_{+}^a}&=&\sin a\ket{\mathbf u_1}+\cos a\ket{\mathbf u_2}, \quad
 \ket{\mathbf u_{-}^a}=\cos a\ket{\mathbf u_1}-\sin a\ket{\mathbf u_2}, \\
 \ket{f_{+}^b}&=&\sin b\ket{f_1}+\cos b\ket{f_2} , \quad
 \ket{f_{-}^b}=\cos b\ket{f_1}-\sin b\ket{f_2}.
 \label{eq:rotations}
\eeqa
We define the projection operators $\Pi_k^U(a)=\proj{\mathbf u_k^a}$ and
$\Pi_l^F(b)=\proj{f_l^b}$ respectively in the $U$ and $F$ vector spaces with $k,l=+,-$.  A complete, lossless four-channel analyzer gives
normalized joint probabilities
\begin{equation}
 P_{kl}(a,b)=
 \bra e\Pi_k^U(a)\otimes\Pi_l^F(b)\ket e
 =\left|\bra{\mathbf u_k^a}\bra{f_l^b}\ket e\right|^2,
 \qquad \sum_{k,l}P_{kl}(a,b)=1.
 \label{eq:joint-intensity}
\end{equation}
These quantities are nonnegative fractions of total intensity.  Calling them probabilities refers to normalized output-channel weights, i.e., the fraction of the output channel over the total intensity. It does not claim
that a classical beam produces an indivisible particle in only one channel.

\subsection{What replaces spatial locality?}
\label{inequality assumptions}

The two vector spaces $U,F$ in Eq.~\eqref{eq:schmidt} need not be spatially separated.
Consequently, Eq.~\eqref{eq:bell-locality} cannot be invoked as relativistic
locality. Instead we consider the joint probabilities for the optical candidate model in the form
\begin{equation}
 P_{kl}(a,b)=\int\dd\lambda\,\rho(\lambda)
 p^U_k(a,\lambda)p^F_l(b,\lambda),
 \label{eq:factorized-response}
\end{equation}
with $\rho(\lambda)\geq0$, $\int\rho(\lambda)\dd\lambda=1$, and the same $\rho$ for all
settings. Here $p^U_k(a,\lambda)$, $p^F_l(b,\lambda)$ represent the probabilities (normalized fractions of the total intensity) in the basis vectors $\ket{\mathbf u_{k}^a}$, $\ket{f_{l}^b}$ respectively. Equation~\eqref{eq:factorized-response} contains three assumptions:
\begin{enumerate}
 \item \emph{measurement independence}: $\rho(\lambda)$ does not depend on
 the analyzer choices;
 \item \emph{subsystem parameter independence}: the response in either
 degree of freedom depends on its own setting but not the setting and outcome of the other independent degree of freedom;
 \item \emph{conditional response factorization}: any remaining correlation
 between the two outputs is mediated by $\lambda$.
\end{enumerate}
The phrase ``independent degrees of freedom'' should be understood here as the ability to address their analyzers independently and as the tensor-factor
structure of the ideal operations.  It must not be interpreted as direct uncorrelation,
$P_{kl}(a,b)=p^U_k(a)p^F_l(b)$, because the nonseparable field generally has correlated outcomes.

A classically nonseparable beam can exhibit correlations between $U$ and $F$. Within the model of Eq.~\eqref{eq:factorized-response}, the two responses are independent for a fixed $\lambda$, but averaging over a common distribution of $\lambda$ allows the resulting joint probabilities to exhibit correlations. The role of $\lambda$ is therefore to allow for correlations within the factorized model. Whether that model can reproduce the full set of joint probabilities across measurement settings is the question addressed by the Wigner–Bell inequality.

For the perfect equal-setting pair, $\kappa_1=\kappa_2$, one has
\begin{equation}
 P_{+-}(\beta,\beta)=P_{-+}(\beta,\beta)=0
 \label{eq:perfect-optical}
\end{equation}
with $a=b=\beta$. Combining it with Eq.~\eqref{eq:factorized-response}, one immediately obtains 
\beqa
 \int\rho(\lambda) p^U_+(\beta,\lambda)\Big[1-p^F_+(\beta,\lambda)\Big]\dd\lambda=0, \quad \int\rho(\lambda)\Big[1-p_+^{U}(\beta,\lambda)\Big]p^F_+(\beta,\lambda) \dd\lambda=0.
\eeqa
Since these two conditions are valid for all $\lambda$, it follows directly $p_+^{U}(\beta,\lambda)=p^F_+(\beta,\lambda)$, a result of the perfect equal-setting correlation. We define $x\equiv p^U_+(\alpha,\lambda)$, $y\equiv p_+^{U}(\beta,\lambda)=p^F_+(\beta,\lambda)$, and
$z\equiv p_+^F(\gamma,\lambda)$, with $x,y,z\in[0,1]$ and achieve the identity
\begin{equation}
 xz+(1-x)y-yz=xz(1-y)+(1-x)y(1-z)\geq0.
 \label{eq:pointwise}
\end{equation}

Averaging the above relation over the hidden variable $\lambda$ with
the normalized distribution $\rho(\lambda)$ gives
\begin{equation}
\int \dd\lambda\rho(\lambda)xz
+
\int \dd\lambda\rho(\lambda)(1-x)y
\geq
\int \dd\lambda\rho(\lambda)yz.
\label{eq}
\end{equation}
Using the factorized expressions for the joint probabilities in
Eq.~\eqref{eq:factorized-response}, the three integrals can be identified
directly with the corresponding experimentally accessible pairwise
probabilities. Equation~\eqref{eq} therefore reduces to the exact Wigner-Bell inequality given in Eq.~\eqref{eq:wigner-bound}.

\begin{figure}[b]
 \centering
 \includegraphics[width=\textwidth]{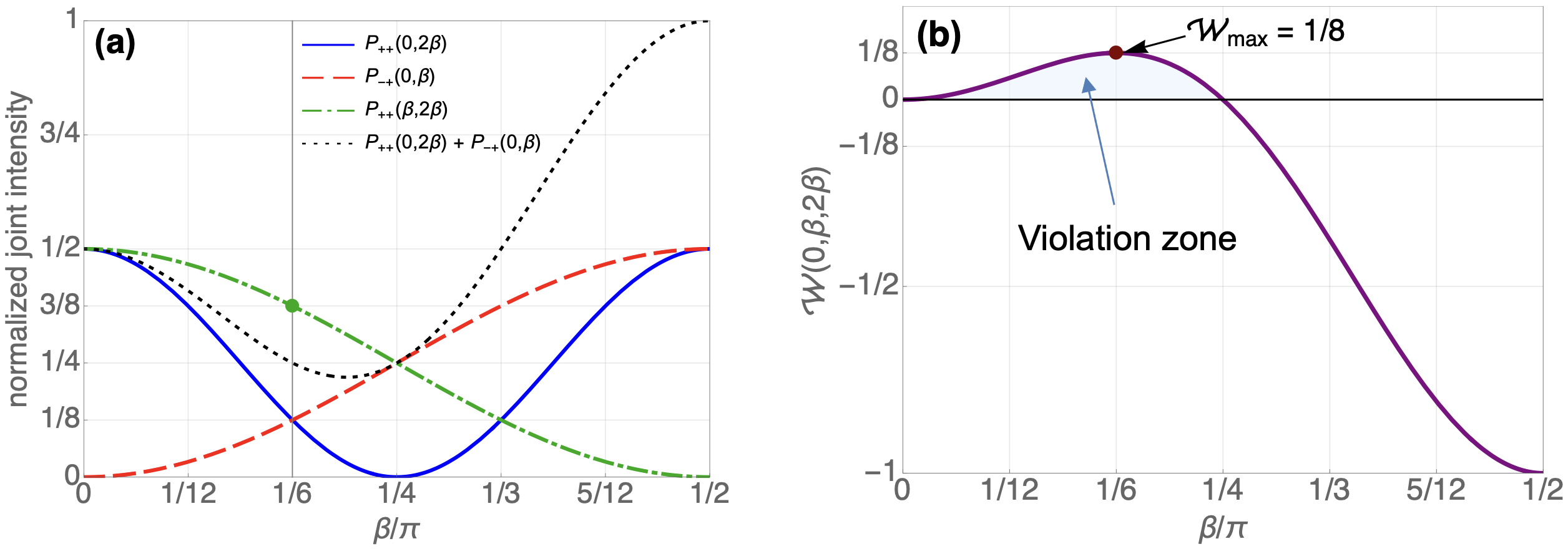}
 \caption{Wigner--Bell violation of the perfect equal-Schmidt field \eqref{eq:equal-schmidt} for
 $\alpha=0$ and $\gamma=2\beta$. (a) The three joint probabilities $P_{++}(\alpha,\gamma), P_{-+}(\alpha,\beta), P_{++}(\beta,\gamma)$ and
 the sum of the first two terms. At $\beta=\pi/6$, the two terms
 on the left are each $1/8$, while the right side is $3/8$. (b) The Wigner excess $\WU$, as function of $\beta$, is positive in the violation region, $0<\beta<\pi/4$, and reaches maximum $1/8$ at $\beta=\pi/6$.}
 \label{fig:violation}
\end{figure}

\section{Violation by a nonseparable classical field}
\label{sec:violation}
To analyze the violation of Wigner-Bell inequality \eqref{eq:wigner-bound}, we consider the perfect equal-setting correlation optical field
\begin{equation}
 \ket{e_+}=\frac{\ket{\mathbf u_1}\ket{f_1}+\ket{\mathbf u_2}\ket{f_2}}{\sqrt{2}}.
 \label{eq:equal-schmidt}
\end{equation}
From the Malus' law, the joint probabilities in Eq.~\eqref{eq:joint-intensity} give
\begin{equation}
 P_{++}(a,b)=P_{--}(a,b)=\frac12\cos^2(a-b),\qquad
 P_{+-}(a,b)=P_{-+}(a,b)=\frac12\sin^2(a-b).
 \label{eq:malus-joint}
\end{equation}
Therefore, the corresponding relevant joint probabilities in Eq.~\eqref{eq:wigner-bound} are obtained straightforwardly as \cite{Eberly2002}
\begin{align}
 P_{++}(\alpha,\gamma)=\tfrac12\cos^2(\alpha-\gamma), \quad
 P_{-+}(\alpha,\beta)=\tfrac12\sin^2(\alpha-\beta), \quad
 P_{++}(\beta,\gamma)=\tfrac12\cos^2(\beta-\gamma).
 \label{eq:three-probabilities}
\end{align}
The behaviors of these joint probabilities are illustrated in detail in Fig.~\ref{fig:violation} (a) when taking $\alpha=0$ and $\gamma=2\beta$. Under this parameter choice, the Wigner-Bell inequality becomes
\begin{equation}
\cos^2(2\beta)\geq\cos(2\beta).
\label{eq:trig-inequality}
\end{equation}

Apparently, this inequality is violated precisely when
\begin{equation}
 0<\cos(2\beta)<1,
 \label{eq:violation-range}
\end{equation}
which is equivalent to $0<|\beta -n \pi| <\pi/4$ with $n$ as an integer. If the polarization angle is defined modulo $\pi$, as is natural for a linear polarizer, one only needs to analyze one fundamental continuous violation interval, e.g., $0<\beta<\pi/4$. Equality of \eqref{eq:trig-inequality} holds when $\cos(2\beta)=0$ or $1$, and the inequality is satisfied for
$\cos(2\beta)<0$.

\begin{figure}[h]
 \centering
\includegraphics[width=0.6\textwidth]{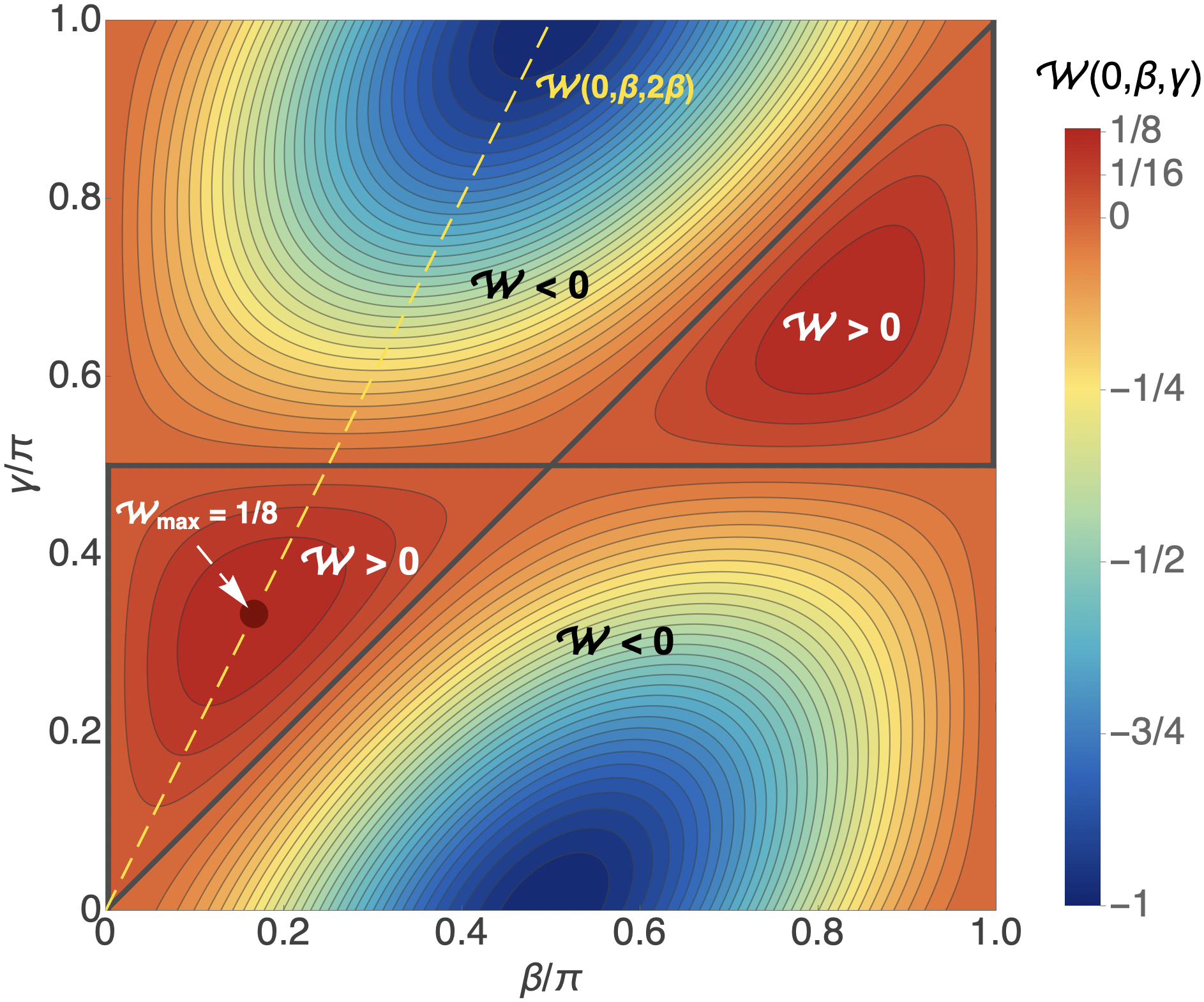}
 \caption{Wigner violation landscape for $\alpha =0$, i.e., the Wigner excess $\WU(\alpha = 0, \beta,\gamma) = \tfrac12 [\cos^2(\beta-\gamma)-\cos^2\gamma-
 \sin^2\beta]$. The gray contours mark $\WU=0$. The positive regions ($\WU>0$) violate the ideal Wigner bound for $0<\beta<\gamma<\pi/2$ and $\pi/2<\gamma<\beta<\pi$ . The dashed line identifies the symmetric family
 $\gamma=2\beta$ same as the line $\WU(\alpha=0,\beta,2\beta)$ in Fig.~\ref{fig:violation} (b), and the marked point is its maximum.}
 \label{fig:landscape}
\end{figure}

To further evaluate the violation of the Wigner-Bell inequality, we define the Wigner excess
\begin{equation}
 \WU(\alpha,\beta,\gamma)=P_{++}(\beta,\gamma)
 -P_{++}(\alpha,\gamma)-P_{-+}(\alpha,\beta),
 \label{eq:Wigner_excess}
\end{equation}
and violation happens when $ \WU(\alpha,\beta,\gamma)>0$.
Along the symmetric family,
\begin{equation}
 \WU(0,\beta,2\beta)=\frac12
 \left[\cos(2\beta)-\cos^2(2\beta)\right].
 \label{eq:W-symmetric}
\end{equation}
The maximum of the Wigner excess is $\WU_{\rm max}=1/8$ at $\cos(2\beta)=1/2$ with $ (\alpha,\beta,\gamma)=(0,\pi/6,\pi/3)$, 
which gives $
 P_{++}(0,\pi/3)=\frac18,\quad
 P_{-+}(0,\pi/6)=\frac18,\quad
 P_{++}(\pi/6,\pi/3)=\frac38$, and hence the inequality becomes $1/4\ge 3/8$, which is apparently violated. Fig~\ref{fig:violation}(b) shows explicitly the Wigner excess $\WU$ as a function of $\beta$ for the symmetric setting $\alpha=0$ and $\gamma=2\beta$. The region in which $\WU>0$ corresponds directly to a violation of the Wigner-Bell inequality.

Importantly, the violation is not an isolated consequence of this symmetric choice of measurement settings. More generally, fixing $\alpha=0$ while
allowing $\beta$ and $\gamma$ to vary independently reveals an extended region of parameter space in which the Wigner--Bell inequality is violated.
As shown in Fig.~\ref{fig:landscape}, the Wigner excess $\WU(0, \beta, \gamma)$ is positive throughout the region enclosed by the
$\WU=0$ triangle boundary in gray solid lines. Thus, the symmetric choice $\gamma=2\beta$ represents
only a one-dimensional cut through a broader two-dimensional violation
regime, rather than a special isolated condition for observing the Wigner-Bell violation.

\section{Extended Wigner--Bell inequality with imperfect correlations}

The ideal Wigner--Bell inequality assumes perfect equal-setting
correlations, resulting Eq.~\eqref{eq:perfect-optical}. In real experiments, however, imperfect mode
preparation, detector noise, alignment errors, etc., generally produce
small but nonzero equal-setting mismatch probabilities. We therefore
derive a robust form of the inequality \eqref{eq:wigner-bound} that remains valid without assuming perfect equal-setting correlations.

For these realistic experimental conditions, the two Schmidt coefficients are in general not equal, i.e., $\kappa_1\neq \kappa_2$ in Eq.~\eqref{eq:schmidt}. Then one has the nonvanishig joint probabilities
\begin{equation}
 P_{+-}(\beta,\beta)=P_{-+}(\beta,\beta)=(\kappa_1-\kappa_2)^2\sin^2\beta\cos^2\beta\neq0.
 \label{eq:imperfect-optical}
\end{equation}
Therefore, in general
$p_+^{U}(\beta,\lambda)\neq p^F_+(\beta,\lambda) \in [0,1]$.

Now we define $u=p_+^U(\beta,\lambda)$, $f=p_+^F(\beta,\lambda)$, and combine with the previous definition $x\equiv p^U_+(\alpha,\lambda)$ and
$z\equiv p_+^F(\gamma,\lambda)$, we construct a difference
\begin{align}
  \mathcal{D}(\lambda)
 &=
 xz+(1-x)f+u(1-f)-uz  \nonumber \\
 &=
 (1-x)\bigl[f(1-u)+u(1-z)\bigr]
+x\bigl[z(1-u)+u(1-f)\bigr]\ge 0,
 \label{eq:imperfect-positive-decomposition}
\end{align}
where we have used the fact that $x,u,f,z\in[0,1]$. As a result one has
\begin{equation}
 xz+(1-x)f+u(1-f)\geq uz
 \label{eq:imperfect-pointwise-bound}
\end{equation}
for every $\lambda$. Multiplying Eq.~\eqref{eq:imperfect-pointwise-bound} by
$\rho(\lambda)$ and integrating over $\lambda$ gives
\begin{align}
 \int\dd\lambda\,\rho(\lambda)xz
 +\int\dd\lambda\,\rho(\lambda)(1-x)f +\int\dd\lambda\,\rho(\lambda)u(1-f)
 \geq
 \int\dd\lambda\,\rho(\lambda)uz.
 \label{eq:imperfect-integrated-bound}
\end{align}

From the joint probabilities in Eq.~\eqref{eq:factorized-response}, the above inequality immediately becomes the experimentally robust extended Wigner--Bell inequality
\begin{equation}
 P_{++}(\alpha,\gamma)
 +P_{-+}(\alpha,\beta)
 +P_{+-}(\beta,\beta)
 \geq
 P_{++}(\beta,\gamma).
 \label{eq:robust-wigner-bound}
\end{equation} 

Consequently, the Wigner excess satisfies the new bound
\begin{equation}
 \mathcal{W}(\alpha,\beta,\gamma)
 \leq P_{+-}(\beta,\beta).
 \label{eq:robust-wigner-excess}
\end{equation}
Thus, when the equal-setting correlation is imperfect, a positive
value of $\mathcal{W}$ alone is not sufficient to demonstrate a
violation. An experimentally significant violation requires
\begin{equation}
 \mathcal{W}(\alpha,\beta,\gamma)
 >
 P_{+-}(\beta,\beta),
 \label{eq:experimental-violation-condition}
\end{equation}
with the difference remaining positive after statistical and
systematic uncertainties are included, where all the joint probabilities are obtained from \eqref{eq:malus-joint} obeying the Malus law.

\section{Proposed optical test} \label{sec:experiment} 
\begin{figure}[t]
 \centering
\includegraphics[width=0.8\textwidth]{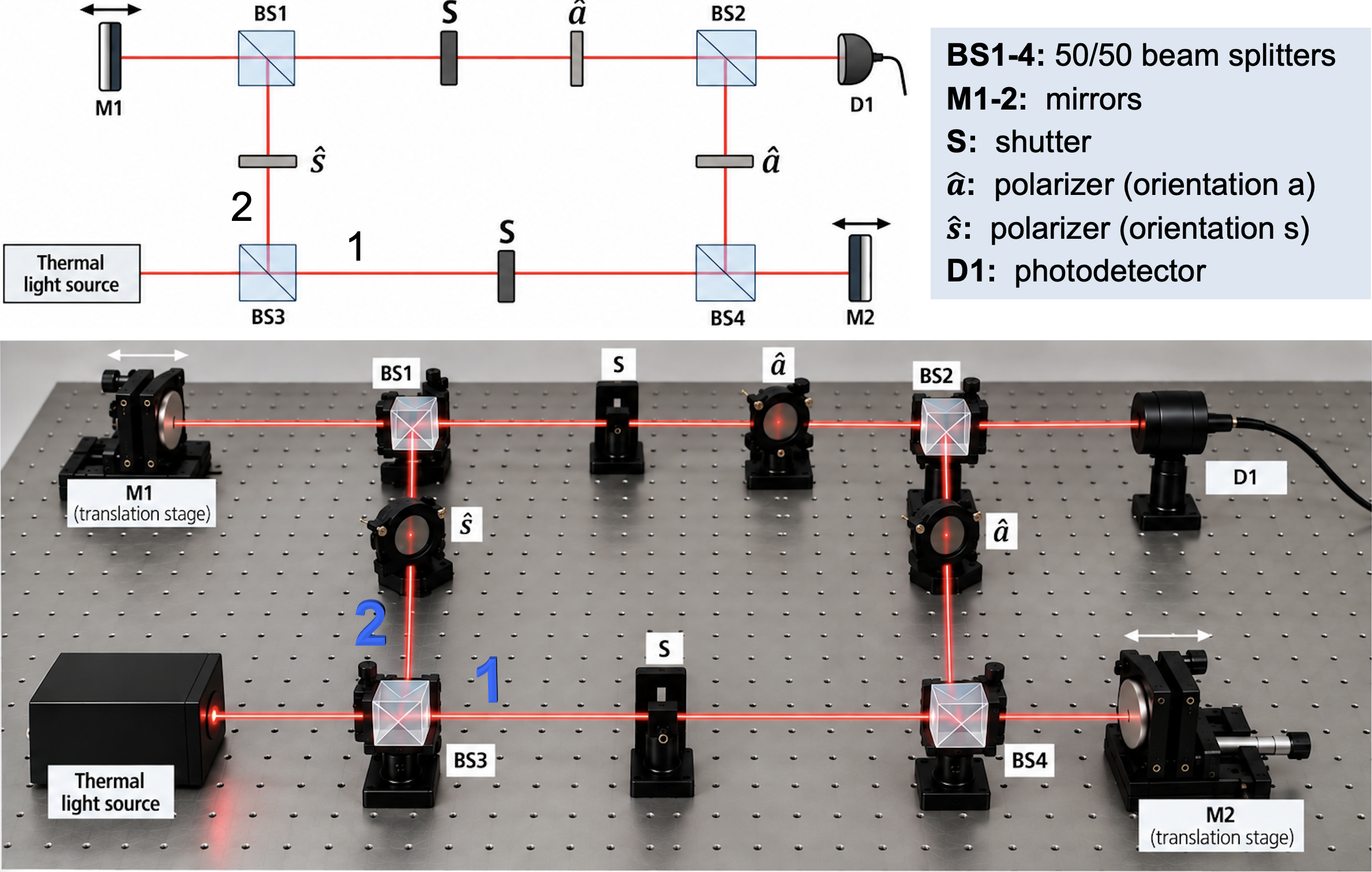}
\caption{Proposed interferometric test of the original and extended Wigner--Bell inequality with
classically nonseparable light. The upper panel shows the optical schematic,
and the lower panel illustrates a corresponding laboratory realization.
A thermal-light field with two nonseparable degrees of freedom is divided at
BS3 into two paths. In path~1, polarizer $\hat{a}$ projects the polarization
degree of freedom onto $\ket{\mathbf u_k^a}$, producing the field
$\ket{\mathbf E_k^a}$. In path~2, polarizer $\hat{s}$ performs the stripping
projection that selects the desired function-space component
$\ket{f_l^b}$, followed by polarizer $\hat{a}$ to prepare the auxiliary field
with the same polarization as the corresponding component in path~1.
The two fields are recombined at the 50/50 beam splitter BS2, and their
interference intensity is recorded by detector D1. Shutters $S$ selectively
block the two paths for the required individual intensity measurements, while
mirrors M1 and M2 on translation stages control the relative optical path
length and interference condition.}
 \label{fig:experiment}
\end{figure}

We now propose an optical experiment for testing the Wigner-Bell inequality in Eq.~\eqref{eq:wigner-bound} and its generalized extension \eqref{eq:robust-wigner-bound}. The input is a classically nonseparable thermal-light field of the Schmidt form given in Eq.~\eqref{eq:schmidt}. Such a field may be generated, for example, from a laser diode operated below threshold, thermal radiation, or other suitably prepared incoherent sources. Figure~\ref{fig:experiment} illustrates the proposed realization, which is adapted from the classical-light CHSH experiment of Ref.~\cite{Qian2015}. The input field is divided into two paths. Path~1 performs a polarization projection onto $\ket{\mathbf u_k^a}$ via polarizer $\hat{a}$, while path~2 prepares an auxiliary field containing a selected function-space component $\ket{f_l^b}$ via polarizers $\hat{s}$ and $\hat{a}$~\cite{Qian2015}. Interference between the two paths then allows the coefficients associated with corresponding components to be determined from directly measurable intensities. In this way, the joint probabilities $P_{kl}(a,b)$ appearing in Eq.~\eqref{eq:joint-intensity} can be reconstructed without directly resolving the function-space mode. The effective joint projection is implemented in two stages. First state involves the projection of the input field onto the polarization state $\ket{\mathbf u_k^a}$ in path~1. A polarizer oriented at the corresponding angle $a$ produces
\begin{equation} 
\ket{\mathbf E_k^a} =\bra{\mathbf u_k^a}\mathbf{E}\ra= \sqrt{I_k^a}\, \ket{\mathbf u_k^a} \left( A_{k+}\ket{f_+^b} + A_{k-}\ket{f_-^b} \right), 
\label{eq:path1-field} 
\end{equation} 
where $I_k^a$ is the intensity after the polarization projection and the coefficients are normalized such that $|A_{k+}|^2+|A_{k-}|^2=1$. The corresponding joint probability is therefore 
\begin{equation} P_{kl}(a,b) = \frac{I_k^a}{I}|A_{kl}|^2, 
\label{eq:Pkl-Akl} 
\end{equation} 
where $I$ is the total input intensity. Although $I$ and $I_k^a$ can be measured directly, the coefficients $A_{kl}$ are not obtained from the polarization projection alone. 

The second stage determines $|A_{kl}|$ thus the joint probability $P_{kl}(a,b)$ interferometrically. Path~2 implements a stripping operation that prepares an auxiliary field of the form 
\begin{equation} 
\ket{\mathbf E_{kl}^{ab}} = \sqrt{I_{kl}^{ab}}\, \ket{\mathbf u_k^a}\ket{f_l^b}. 
\label{eq:auxiliary-field} 
\end{equation} 
This field can be generated using two successive polarization projections, denoted by $\hat{s}$ and $\hat{a}$. The first polarizer is set to an angle $s$ satisfying 
\begin{equation} 
\tan s = \frac{\kappa_2}{\kappa_1}\tan b \quad \text{and} \quad \tan s = \frac{\kappa_1}{\kappa_2}\tan b,
\label{eq:stripping-angle} 
\end{equation} 
for $l=+$ and $l=-$ respectively. With this choice, projection onto $\ket{\mathbf u_k^s}$ selects the corresponding function-space component $\ket{f_k^b}$ and produces $\ket{\mathbf E_{kl}^{sb}} = \sqrt{I_{kl}^{sb}} \ket{\mathbf u_k^s}\ket{f_l^b}$. A subsequent polarization projection onto $\ket{\mathbf u_k^a}$ then gives precisely the polarization and function-space structure, as required in \eqref{eq:auxiliary-field}, for interference with the corresponding component of the path~1 field. 

The beams from the two paths are then combined at a 50/50 beam splitter BS2. Choosing the difference output port gives \begin{equation} 
\ket{\mathbf E_k^T} = \frac{ \ket{\mathbf E_k^a} - \ket{\mathbf E_{k+}^{ab}} }{\sqrt{2}}, 
\label{eq:interference-field} \end{equation} 
with measured intensity $I_k^T$. For mutually phase-referenced fields, the interference term determines the amplitude of the selected component, i.e.,
\beq
2I_k^T =I_{k+}^{ab}+I_k^a+\sqrt{I_{k+}^{ab}I_k^a}|A_{k+}|(e^{i\Delta}+e^{-i\Delta}),
\eeq
where $\Delta$ is the combination of the phase in the complex amplitude $A_{k+}$ and any external adjustable relative phase of the two arms. At perfect constructive interference, the total intensity reaches maximum
\beq
2I_k^T =I_{k+}^{ab}+I_k^a+2\sqrt{I_{k+}^{ab}I_k^a}|A_{k+}|.
\eeq
Consequently, the joint probabilities can consequently be reconstructed as 
\begin{align} 
P_{k+}(a,b) &= \frac{ \left( I_{k+}^{ab}+I_k^a-2I_k^T \right)^2 }{ 4I I_{k+}^{ab} }, \label{eq:P-kplus-exp} \\ 
P_{k-}(a,b) &= \frac{I_k^a}{I} - \frac{ \left( I_{k+}^{ab}+I_k^a-2I_k^T \right)^2 }{ 4I I_{k+}^{ab} }. 
\label{eq:P-kminus-exp} 
\end{align} 
The second relation follows directly from the marginal-probability identity $P_{k+}(a,b) + P_{k-}(a,b) = I_k^a/I$. Thus, sequential measurements of intensities $I, I_k^a, I_{k+}^{ab}, I_k^T$ are sufficient to recover the joint probabilities to test the original and extended Wigner--Bell inequality. Here, optimal constructive interference is required when measuring $I_k^T$. Normalization of intensity measurements for each run is achieved with complementary intensity registration for the other exit path of BS2.

For the specific inequality considered here \eqref{eq:wigner-bound}, the three required joint probabilities are obtained from 
\beqa
P_{++}(\alpha,\gamma) &=&\frac{\left( I_{++}^{\alpha\gamma}+I_{+}^{\alpha}-2I_{+}^{T}\right) ^{2}}{4II_{++}^{\alpha\gamma}}, \\
P_{-+}(\alpha,\beta) &=&\frac{\left( I_{-+}^{\alpha\beta}+I_{-}^{\alpha}-2I_{-}^{T}\right) ^{2}}{4II_{-+}^{\alpha\beta}}, \\
P_{++}(\beta,\gamma) &=&\frac{\left( I_{++}^{\beta\gamma}+I_{+}^{\beta}-2I_{+}^{T}\right) ^{2}}{4II_{++}^{\beta\gamma}}.
\label{eq:three-measured-probabilities}
\eeqa
For the generalized extended inequality \eqref{eq:robust-wigner-bound}, an additional joint probability will be required
\begin{align} 
P_{+-}(\beta,\beta) &= \frac{I_+^{\beta}}{I} - \frac{ \left( I_{++}^{\beta\beta}+I_+^{\beta}-2I_+^T \right)^2 }{ 4I I_{++}^{\beta\beta} }. 
\end{align} 
Here the superscripts on the interference intensities explicitly identify the corresponding analyzer settings. Varying $\alpha$, $\beta$, and $\gamma$ therefore provides a direct experimental test of the theoretical predictions in Eqs.~\eqref{eq:trig-inequality} and \eqref{eq:W-symmetric}, including the violation regions shown in Figs.~\ref{fig:violation} and \ref{fig:landscape}.

Although polarization is used here for the first degree of freedom, the second two-dimensional space need not have a particular physical realization. It may be encoded in spatial modes, paths, temporal modes, frequency modes, or another pair of calibrated orthogonal optical modes. The essential requirement is a classically nonseparable field with a controllable two-dimensional Schmidt structure and the ability to prepare and interfere the corresponding projected components. This experiment therefore converts the Wigner--Bell test into a set of directly accessible intensity and interference measurements.

\section{Interpretation: what the violation does and does not show}
\label{sec:interpretation}

The field of Eq.~\eqref{eq:equal-schmidt} is a solution within classical wave
optics and its projected intensities follow the usual squared-amplitude rule. A positive value of Eq.~\eqref{eq:Wigner_excess} does not invalidate Maxwell's equations, probability additivity within a fixed
four-channel measurement, or the existence of a classical wave description.
Nor does it demonstrate superluminal influence or spatial Bell nonlocality:
the two analyzed spaces are internal degrees of freedom of one optical field.

The excluded object is more specific.  Under the measured equal-setting
correlation, there is no single nonnegative distribution assigning all
alternative outcomes in Eq.~\eqref{eq:globaldistribution} that both reproduces
the contextual pairwise data and satisfies the {\em setting-independent factorized
response representation}.  In the language of marginal scenarios, the
observed pairwise distributions do not admit the stipulated global extension.
This is closely related to contextuality, although a claim of generalized
noncontextuality should specify operational equivalences and assumptions with
the care emphasized in Ref.~\cite{Bell1966,KochenSpecker1967, Spekkens2005, Li2017Contextuality}.

Malus' law plays a direct calculational role: it supplies the squared trigonometric dependence responsible for Eq.~\eqref{eq:trig-inequality}.
Conceptually, however, Malus' law alone is not a Bell assumption and its
validity is not what the violation challenges.  The tension arises when
Malus-law probabilities collected in different contexts are required to be
marginals of the global model.

\section{Conclusion}
\label{conclusion}
We have developed a Wigner-Bell analysis for classically nonseparable light that makes explicit the probabilistic structure underlying the inequality. Starting from the conventional Bell framework, we showed that the Wigner formulation can be understood as a consistency condition requiring pairwise probabilities obtained in different measurement contexts to arise as marginals of a single nonnegative global joint distribution. For two internal degrees of freedom of a classical optical field, spatial locality is not the relevant physical assumption. Instead, we formulated the corresponding candidate model in terms of measurement independence and subsystem response factorization, together with the equal-setting correlation required by the Wigner construction. These assumptions lead directly to the Wigner-Bell bound \eqref{eq:wigner-bound} in the equal-setting perfect correlation case and the general extended Wigner-Bell bound \eqref{eq:robust-wigner-bound} for arbitrary imperfect correlations in realistic experimental conditions.

For a nonseparable classical field, the relevant probabilities are normalized joint intensities and follow directly from Malus' law \cite{Eberly2002}. We showed, for the perfect equal-Schmidt field, that these probabilities violate the Wigner--Bell bound over a finite and extended range of analyzer settings. For the symmetric choice $\alpha=0$ and $\gamma=2\beta$, the violation occurs for $0<\beta<\pi/4$ and reaches a maximum Wigner excess of $\WU_{\rm max}=1/8$ at $\beta=\pi/6$. More generally, the violation occupies a two-dimensional region of the $(\beta,\gamma)$ parameter space, demonstrating that it is not a consequence of a special isolated choice of analyzer settings.

The significance of this violation is importantly different from that of a Bell test performed with spatially separated quantum particles. The two subsystems considered here are internal degrees of freedom of a single classical field, and therefore the violation does not demonstrate spatial nonlocality, superluminal influence, or a breakdown of classical electrodynamics. Rather, it rules out the specific global probability construction underlying the Wigner bound: the Malus-law pairwise statistics obtained in different measurement contexts cannot all be embedded as marginals of a single nonnegative global joint distribution satisfying the stipulated factorized-response assumption, i.e., assumption 3 in section \ref{inequality assumptions}. In this sense, the violation isolates the tension between context-dependent projection statistics and a context-independent global assignment of simultaneous outcomes.

We have also proposed an interferometric realization in which the required joint probabilities can be reconstructed entirely from measurable optical intensities. Because the construction relies on the tensor-product structure of two optical vector spaces rather than on a particular physical encoding, the second degree of freedom may be realized using spatial, temporal, path, frequency, or other orthogonal optical modes. The proposed experiment therefore provides a practical route for directly testing the Wigner probability bound with classical light.

More broadly, this analysis illustrates why the observation of a Bell-type inequality violation and the physical interpretation of that violation must be carefully distinguished. Hilbert-space structure, nonseparability, projection, and squared-amplitude laws already produce correlations in classical wave systems that can be incompatible with particular global joint-probability models. Spatial quantum nonlocality requires additional physical structure that is absent from the present optical system. Classically nonseparable light therefore provides a particularly transparent platform for separating these ingredients and for identifying which features of Bell-type phenomena originate from general wave and probability structure and which are uniquely associated with genuinely quantum systems.

\section*{Acknowledgments}
Many of the ideas developed in this article were inspired by conversations
with Joseph H. Eberly over the years. We are also grateful to Michael Raymer,
Gerd Leuchs, and Girish Agarwal for insightful discussions. We acknowledge
financial support from the National Science
Foundation under Grants No.~PHY-2316878 and No.~PHY-2514953, and the U.S. Army under Contract No.~W15QKN-24-C-0004.

\section*{Data availability}
No data were generated for this study.


\end{document}